\documentclass{llncs}

\usepackage{natbib}
\usepackage{graphicx}

\usepackage{fancyhdr}
\usepackage{lastpage}
 
\begin{document}

\title{A short review and primer on electromyography in human computer interaction applications}
\author{Niklas Ravaja\inst{1,2,3} \and Benjamin Cowley\inst{4,5} \and Jari Torniainen\inst{4}}
\institute{
Helsinki Collegium for Advanced Studies, University of Helsinki, Finland
\and
Helsinki Institute for Information Technology, Aalto University, Finland
\and
School of Business, Aalto University, Finland
\and
Quantitative Employee unit, Finnish Institute of Occupational Health,\\
\email{benjamin.cowley@ttl.fi},\\
POBox 40, Helsinki, 00250, Finland
\and
Cognitive Brain Research Unit, Institute of Behavioural Sciences, University of Helsinki, Finland}

\maketitle              

\begin{abstract}
The application of psychophysiology in human-computer interaction is a growing field with significant potential for future smart personalised systems. Working in this emerging field requires comprehension of an array of physiological signals and analysis techniques. 

Electromyography (EMG) is a useful signal to estimate the emotional context of individuals, because it is relatively robust, and simple to record and analyze. Common uses are to infer emotional valence in response to a stimulus, and to index some symptoms of stress. However, in order to interpret EMG signals, they must be considered alongside data on physical, social and intentional context. Here we present a short review on the application of EMG in human-computer interaction. 

This paper aims to serve as a primer for the novice, enabling rapid familiarisation with the latest core concepts. We put special emphasis on everyday human-computer interface applications to distinguish from the more common clinical or sports uses of psychophysiology.

This paper is an extract from a comprehensive review of the entire field of ambulatory psychophysiology, including 12 similar chapters, plus application guidelines and systematic review. Thus any citation should be made using the following reference:

{\parshape 1 2cm \dimexpr\linewidth-1cm\relax
B. Cowley, M. Filetti, K. Lukander, J. Torniainen, A. Henelius, L. Ahonen, O. Barral, I. Kosunen, T. Valtonen, M. Huotilainen, N. Ravaja, G. Jacucci. \textit{The Psychophysiology Primer: a guide to methods and a broad review with a focus on human-computer interaction}. Foundations and Trends in Human-Computer Interaction, vol. 9, no. 3-4, pp. 150--307, 2016.
\par}


\end{abstract}

\section{Introduction}

Electromyography (EMG) involves the detection of myoelectric potentials by means of surface electrodes. It measures the electrical activity associated with contractions of striated muscles \citep{Tassinary2000}. These muscle contractions may yield a direct index of the physical embodiment of various mental states, including emotions, stress, or fatigue; for example, the contraction of facial muscles underlies some emotional expressions. When one is assessing facial expression in emotion, the advantage of facial EMG measurement over observation (facial expression coding) is that it can sensitively assess hidden facial muscle activity that may not be perceptible by mere observation \citep{Ravaja2004}.

\section{Background}

Facial EMG has been found to be a successful method primarily in discriminating positive emotions from negative ones \citep{Ravaja2004}. That is, facial EMG is a psychophysiological index of hedonic valence, the dimension of emotion that ranges from negative (or unpleasant) to positive (or pleasant). For producing this index, facial EMG activity is usually recorded over three distinct facial muscle areas: the zygomaticus major (the cheek muscle area that activates during smiling), corrugator supercilii (the brow muscle area that activates during frowning), and orbicularis oculi (the periocular muscle area that activates during the so-called `enjoyment smile' \citep{Tassinary2000}).

A large body of evidence shows that the processing of pleasant emotions is associated with increased activity within the zygomaticus major muscle area and that processing of unpleasant emotions evokes higher activity in the corrugator supercilii muscle area during affective imagery \citep{Ravaja2006} and when the subject is presented with acoustic stimuli \citep{Bradley2000}, radio advertisements \citep{BOLLS2001}, emotional still and moving images (of 6 s duration) \citep{Simons1999}, written words \citep{Larsen2003}, textual news messages \citep{Ravaja2006}, and news messages in video format \citep{Ravaja2006}. There is also evidence that the zygomaticus major responds only to positive valence (pleasantness), while the corrugator supercilii responds to both negative and positive valences, in a reciprocal manner \citep{Larsen2003}. Activity in the orbicularis oculi muscle area has been correlated with pleasantness -- in particular, high-arousal positive emotional states \citep{Ravaja2004} -- and activity in this region has also been found to differentiate smiling due to genuine pleasure from `forced' smiling.

\begin{figure}[!t]
   \centering
   \includegraphics[scale=1.0]{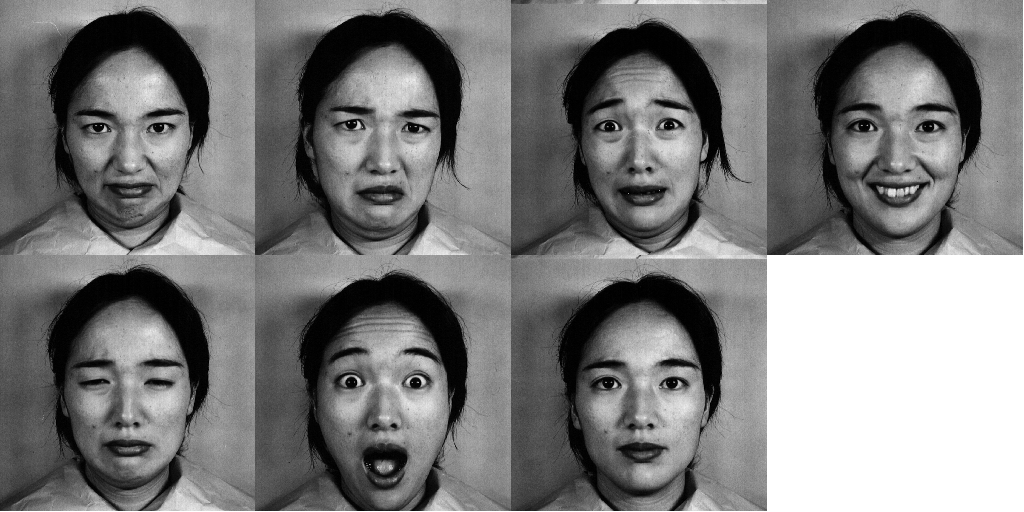}
   \caption{\textbf{Seven actor poses for emotional expressions, from the Japanese Female Facial Expression (JAFFE) database} \citep{Lyons1998}. Clockwise from top left: anger, disgust, fear, happiness, a neutral expression, surprise, and sadness. Although use of posed emotional expressions is a dated technique, such datasets provide a convenient picture of the concepts involved.}
   \label{fig.japfacs}
\end{figure}

When one is interpreting facial EMG measurements, or facial expressions in general, it is important to know whether they provide information on the true emotional state of an individual or are social signals possibly without any connection to emotional experience. That is, there are two competing views. The emotion-expression view is that facial displays express a person's internal emotional state (e.g., \citet{Ekman1994}; see Figure~\ref{fig.japfacs}), whereas the behavioural ecology view holds that facial displays are social signals that communicate behavioural intentions or social motives (and are not `readouts' of an underlying affective state \citep{Fridlund1991}). Although the emotion-expression view is supported by a number of studies (see above), the behavioural ecology view has gained support through studies showing that positive and negative emotional facial displays are augmented in the presence of a real or an imagined audience when one is viewing videotapes, with the effect being independent of a concurrent emotional state \citep{Fridlund1991}. However, all in all, the evidence shows that, in social situations, facial expressions are affected by both emotional and social factors (e.g., \citet{Hess1995}).

In addition, a more straightforward interpretation of EMG signals has been posited, in which facial EMG measured from the temporalis, masseter, and medial and lateral pterygoid muscles can be used to detect bruxism (grinding of the teeth) \citep{Lavigne2008}. Among the possible causes for bruxism while one is awake are stress, anxiety, and hyperactivity. Measurement of EMG from muscles around the neck and shoulder region (such as the upper trapezius and deltoid muscles) can be used to monitor the fatigue of an office worker, for example \citep{Mathiassen1995}. Measurements in the area of the upper trapezius can be used for indexing shoulder--neck load, but these are sensitive to arm motions \citep{Mathiassen1995}.

\section{Methods}

The elementary functional unit of musculature is the motor unit, consisting of a motoneuron and the set of consonant muscle fibres it innervates. Muscles act more or less quickly and precisely in proportion to the innervation ratio -- i.e., the number of muscle fibres per motoneuron -- and this has implications for interpreting the spectral signature of muscle action potentials. For example, such potentials at a given frequency band and power level might correspond to meaningful activity if measured from the orbicularis area but may fall below the threshold required for distinguishing signal from noise if measured from the deltoid. Muscle action potentials propagate rapidly from the motoneuron endplate across muscle fibre, and a small portion of the changing electrical field of (typically multiple) muscle fibres is conducted through the intervening fluid to the skin. Therefore, what EMG measures directly is changing electrical potentials associated with grouped muscle activity, with a possibly very broad frequency range whose characteristics are related to the underlying muscle dimensions.

Facial EMG activity is typically recorded over the above-mentioned muscle areas (on the left side of the face) by means of surface Ag/AgCl electrodes with a contact area 4 mm in diameter (filled with conductive electrode gel \citep{Tassinary2000}). Precise placement of the electrodes in this regard is important, and, for obtaining a good-quality signal, careful preparation of the skin is necessary also, to reduce any impedance between the skin surface and the gel. This involves, for example, rubbing the skin with a gauze pad and cleansing the site with either alcohol or soap and water.

The raw EMG signal is amplified, and frequencies below 30~Hz and above 500~Hz are filtered out (in some conditions, slightly different cut-off frequencies may be applicable \citep{Blumenthal2005,Tassinary2000}). In light of the Nyquist--Shannon sampling theorem, it is important that the sampling rate be at least 1000~Hz (twice the highest frequency featuring in the signal). In the next stage of processing, the signal is rectified (conversion to absolute values is performed) and integrated or smoothed. There are various procedures for analogue and digital processing of the facial EMG signal (see \citet{Blumenthal2005}), but one commonly applied procedure is smoothing, which involves passing the rectified EMG signal to a low-pass filter, using a digital routine. Signal amplitude is thus interpreted as muscle activation; therefore, the final metrics for EMG are simply first-order statistics of the signal, such as maximum amplitude. 

\begin{figure}[!ht]
   \centering
   \includegraphics{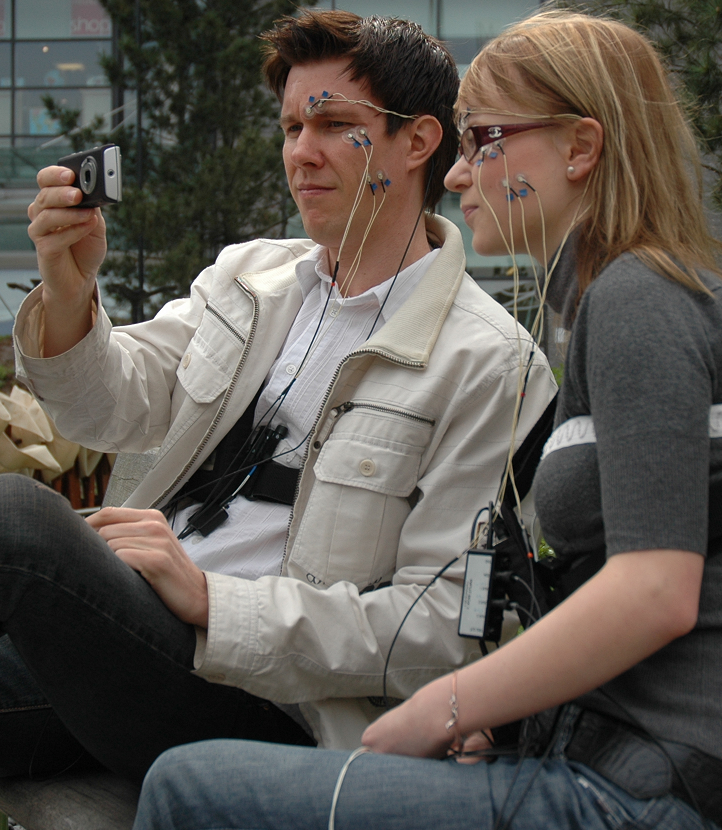}
   \caption{Facial EMG measurement over the brow, periocular, and cheek muscle areas via a mobile physiological data collection system in a real-world context.}
   \label{fig.emg}
\end{figure}

Nowadays, in addition to laboratory devices, there are ambulatory psychophysiological data collection systems, such as the Varioport-ARM (Becker Meditec, Karlsruhe, Germany), that enable the recording of facial EMG in real-world situations (see Figure~\ref{fig.emg}). However, the feasibility of real-world facial EMG recordings is limited by the need to attach electrodes to the face.

\section{Applications}

Facial EMG measurements can be applied to investigate many research questions in the area of HCI. However, as is the case for all psychophysiological measures, the interpretation of facial EMG is highly dependent on the context and research paradigm \citep{Ravaja2004}. It seems very likely that facial EMG provides an index of internal emotional state, especially in contexts wherein little social interaction is involved, such as viewing of emotion-eliciting material (e.g., emotional images and news items) on a computer screen. However, emotional expressions are affected also by display rules (learned rules dictating the management of emotional expressions on the basis of social circumstances, as described by \citet{Zaalberg2004}). Obviously, in some social situations (e.g., when interacting with a person of higher social status), people may even express the emotion opposite what they are feeling (e.g., smiling instead of showing anger). Clear evidence has recently emerged that display rules may influence facial expressions, and also facial EMG activity, when the subject is interacting with a virtual character \citep{ravaja2015virtual}. 
Accordingly, it seems clear that facial EMG can be used to assess emotion-bearing facial expressions but not a human's inner emotional state.

Expressing the opposite emotion from what one is actually feeling may also be done for purposes of emotional coping. For example, there is a recurrent finding that failure in a digital game (e.g., the death of the player's character in a first-person shooter game) elicits increased zygomatic and orbicularis oculi activity and decreased corrugator activity \citep{VandenHoogen2012}. That is, people tend to smile even though the game event is likely to have elicited negative rather than positive emotion. Therefore, it is important to understand that facial EMG does not index inner emotional state in connection with such a game event, even though it may be non-problematically related to experienced emotions in connection with other types of game events. Also, \citet{Ravaja2006opponent} also examined the influence of opponent type on facial EMG activity in playing of digital games. With both co-located and non-co-located players, they found that self-reported pleasure and zygomatic and orbicularis oculi EMG activity increased and corrugator EMG activity decreased in the following order: playing against a computer --- playing against a stranger --- playing against a friend. In this study, facial EMG activity was paralleled by self-reported emotional valence.

Recently, \citet{Salminen2013} studied the effects of computer-mediated cues of group emotion on facial EMG activity when the members of a non-co-located group performed knowledge-work tasks. Negative cues of group emotion (depressed or nervous/stressed) displayed in textual form on a Web page elicited lower self-reported pleasure, less perceived confidence in other group members, and higher corrugator supercilii EMG activity than did cues of positive group emotions (pleasantly excited or pleasantly relaxed). This finding was apparently due to emotional contagion. Thus, facial EMG activity appears to measure inner emotional state in the context of distributed knowledge work, at least when social communication with facial expressions is not relevant (cf. video calls). However, it was also found that planning (creative) tasks elicited lower corrugator EMG activity when compared with routine tasks. Given that routine tasks (checking the grammar of text excerpts, for instance) require effort and that attention and corrugator activity may increase with attentional effort \citep{Cohen1992}, differences in corrugator EMG activity between task types may be explained by attentional requirements. Accordingly, it should be recognised that a given psychophysiological parameter, such as corrugator EMG activity, may index different psychological processes in connection with different factors in a factorial-design experiment. This underscores the importance of understanding differences between experimental conditions in terms of what facial-EMG-related psychological processes they may evoke. 

The EMG technique has been applied to other locations on the body in research on ergonomics and prosthetics. For example, EMG measurements from the deltoid have been used to study the effect of posture on performance in work at a computer \citep{Straker1997}. Wearable EMG trousers can register the activity of leg muscles during standing and walking \citep{Tikkanen2013}, and EMG can also be used as a control signal for prosthetic limbs. Additionally, it can act as an input signal for controlling a computer, whether in everyday applications or as a prosthetic control signal for industrial limb augmentation or rehabilitation. In everyday applications, multi-channel EMG is often used in combination with an accelerometer for gesture recognition in connection with various commands \citep{Zhang2009}. In the case of hand and arm prostheses, the EMG electrodes are usually fixed to one of the antagonist muscle pairs \citep{Zecca2002}.

\section{Conclusions}

It seems apparent that facial EMG can be a valuable measure in the area of HCI. As with all psychophysiological signals, interpretation of the data may pose challenges, though, and only under certain circumstances can facial EMG activity be used to index an individual's inner emotional state. These facts do not, however, diminish the value of facial EMG measurements, and there may be just as much value in obtaining information on emotional facial expressions determined by display rules in computer-mediated communication.

\bibliographystyle{plainnat}
\bibliography{ch4_emg_bib}

\end{document}